\documentclass[onecolumn]{revtex4}
\usepackage{amsmath,amssymb}
\usepackage{graphicx}
\usepackage{dcolumn}
\usepackage{epsfig,color}
\usepackage{bm}
\begin{document}

\title{Open flavor strong decays\thanks{Invited talk presented at NSTAR:Nucleon Resonances: From Photoproduction to High Photon Virtualities}}

\author{H. Garc{\'{\i}}a-Tecocoatzi}
\affiliation{Instituto de Ciencias Nucleares, Universidad Nacional Aut\'onoma de M\'exico, 04510 M\'exico DF, M\'exico}
\affiliation{INFN, Sezione di Genova, via Dodecaneso 33, 16146 Genova, Italy}
\author{R. Bijker}
\affiliation{Instituto de Ciencias Nucleares, Universidad Nacional Aut\'onoma de M\'exico, 04510 M\'exico DF, M\'exico}
\author{J. Ferretti}
\affiliation{Dipartimento di Fisica and INFN, `Sapienza' Universit\`a di Roma, P.le Aldo Moro 5, I-00185 Roma, Italy}
\author{G. Galat\`a}
\affiliation{Instituto de Ciencias Nucleares, Universidad Nacional Aut\'onoma de M\'exico, 04510 M\'exico DF, M\'exico}
\author{E. Santopinto}
\affiliation{INFN, Sezione di Genova, via Dodecaneso 33, 16146 Genova, Italy}

\begin{abstract}
In this contribution, we discuss the results of a QM calculation of the open-flavor strong decays of **** light nucleon resonances. 
These are the results of a recent calculation, where we used a modified $^3P_0$ model for the amplitudes and the U(7) algebraic model and the Hypercentral Quark Model to predict the baryon spectrum. The decay amplitudes are compared with the existing experimental data.
\end{abstract}

\maketitle

\section{Introduction}
\label{intro}
The study of the strong decay processes of baryons is still considered a challenge within theoretical and experimental hadronic physics. At the moment, the number of known light-quark mesons is much larger than the number of known baryon resonances \cite{Nakamura:2010zzi}. However, it is known that the baryon spectrum is  much more complex than the meson one.
The difficulty lies in identifying those high-lying baryon resonances that are only weakly coupled to the $N \pi$ channel \cite{Capstick:1992uc,Capstick:1992th} and thus cannot be seen in elastic $N \pi$ scattering experiments. 

Regarding the strong decays of baryon resonances, no satisfactory description has yet been achieved. We could list several problems as, for example, the QCD mechanism behind the OZI-allowed strong decays \cite{Okubo:1963fa}, which is still not clear. 
Theoretical calculations of baryon strong, electromagnetic and weak decays  still help the experimentalists in their search of those resonances that are still unknown,
even if interesting results were provided by CB-ELSA \cite{Crede:2003ax}, TAPS \cite{Krusche:1995nv}, GRAAL \cite{Renard:2000iv}, SAPHIR \cite{Tran:1998qw} and CLAS \cite{Dugger:2002ft}.

Several phenomenological models have been developed in order to carry out strong decay studies, including pair-creation models \cite{Micu,LeYaouanc,Eichten:1978tg,Alcock:1983gb,Dosch:1986dp,Kokoski:1985is,Roberts:1992,Ackleh:1996yt}, elementary meson emission models \cite{Becchi,Faiman:1968js,Koniuk:1979vy,Godfrey:1985xj,Sartor:1986qr,Bijker:1996tr} and effective Lagrangian approaches (for example, see Ref. \cite{Colangelo:2012xi}). 
A few years after the introduction of  the $^3P_0$ model \cite{Micu}, Le Yaouanc $et$ $al.$ used it to compute meson and baryon open flavor strong decays \cite{LeYaouanc} and also evaluated the strong decay widths of charmonium states \cite{LeYaouanc02}.
The $^3P_0$ model, extensively applied to the decays of light mesons and baryons \cite{Blundell:1995ev}, has been recently applied to heavy meson strong decays, in the charmonium \cite{Barnes:2005pb,Ferretti:2013faa,Ferretti:2014xqa}, bottomonium \cite{Ferretti:2014xqa,Ferretti:2013vua} and open charm \cite{Close:2005se,Segovia:2012cd} sectors. 
In the 90's, Capstick and Roberts calculated the $N \pi$ and the strange decays of nonstrange baryons \cite{Capstick:1992th}, using relativized wave functions for the baryons and mesons.  

Recently, we have computed the decay widths of baryon resonances into baryon-pseudoscalar meson pairs \cite{strong2015} within a modified $^3P_0$ model, using two different models for the mass spectrum: the $U(7)$ algebraic model \cite{Bijker:1994yr-first,Bijker:1994yr}, by Bijker, Iachello and Leviatan, and the hypercentral model (hQM) \cite{pl}, developed by Giannini and Santopinto. The widths have been computed with harmonic oscillator wave functions.
In this contribution, we discuss our main results for the two-body strong decay widths of **** nucleon resonances. 

\section{U(7) model for baryons}
\label{U7}
\label{U(7) algebraic model}
The baryon spectrum is computed by means of algebraic methods introduced by Bijker, Iachello and Leviatan \cite{Bijker:1994yr-first,Bijker:1994yr}. The algebraic structure of the model consists in combining the symmetry of the internal spin-flavor-color part, $SU_{\rm sf}(6) \otimes SU_{\rm c}(3)$, with that of the spatial part, $U(7)$ into 
\begin{equation} 
U(7) \otimes SU_{\rm sf}(6) \otimes SU_{\rm c}(3) ~.
\end{equation}
The $U(7)$ model was introduced \cite{Bijker:1994yr-first} to describe the relative motion of the three constituent parts of the baryon. The general idea is to introduce a so-called spectrum generating algebra $U(k+1)$ for quantum systems characterized by $k$ degrees of freedom. For baryons there are the $k=6$ relevant degrees of freedom of the two relative Jacobi vectors
$
	\vec{\rho} = \frac{1}{\sqrt{2}} (\vec{r}_1-\vec{r}_2) $ and
$
	\vec{\lambda} = \frac{1}{\sqrt{6}} (\vec{r}_1+\vec{r}_2-2\vec{r}_3) $
and their canonically conjugate momenta, $\vec{P_\rho} = \frac{1}{\sqrt{2}} (\vec{p}_1-\vec{p}_2)$ and $\vec{P_\lambda} = \frac{1}{\sqrt{6}} (\vec{p}_1+\vec{p}_2-2\vec{p}_3)$.
The $U(7)$ model is based on a bosonic quantization 
which consists in introducing two vector boson operators $b^{\dagger}_{\rho}$ and $b^{\dagger}_{\lambda}$ 
associated to the Jacobi vectors, and an additional auxiliary scalar boson, $s^{\dagger}$.  
The scalar boson does not represent an independent degree of freedom, but is added 
under the restriction that the total number of bosons $N$ is conserved. The model space consists 
of harmonic oscillator shells with $n=0,1,\ldots,N$.  

The baryon mass formula is written as the sum of three terms
\begin{equation}
	\hat{M}^2 = M_0^2 + \hat{M}^2_{\rm space} + \hat{M}^2_{\rm sf}  \mbox{ },  
\end{equation}
where $M_0^2$ is a constant, $\hat{M}^2_{\rm space}$ is a function of the spatial degrees of freedom and $\hat{M}^2_{\rm sf}$ depends on the internal ones. The spin-flavor part is treated in the same way as in Ref.~\cite{Bijker:1994yr} in terms of a generalized G\"ursey-Radicati formula \cite{Gursey:1992dc}, which in turn is a generalization of the Gell-Mann-Okubo mass formula \cite{Ne'eman,Gell-Mann:1962xb}.

Since the space-spin-flavor wave function is symmetric under the permutation group $S_3$ of three identical constituents, the permutation symmetry of the spatial wave function has to be the same as that of the spin-flavor part. Thus, the spatial part of the mass operator $\hat{M}^2_{\rm space}$ has to be invariant under the $S_3$ permutation symmetry. 
The dependence of the mass spectrum on the spatial degrees of freedom is given by:
\begin{equation}
\hat{M}^2_{\rm space} = \hat{M}^2_{\rm vib} + \hat{M}^2_{\rm rot}  \mbox{ }.  \label{eqn:U(7)02}
\end{equation}

The baryon wave functions are denoted in the standard form as  
\begin{equation}
\left| \, ^{2S+1}\mbox{dim}\{SU_f(3)\}_J \,
[\mbox{dim}\{SU_{sf}(6)\},L_i^P] \, \right> ~, \label{wf}
\end{equation}
where $S$ and $J$ are the spin and total angular momentum $\vec{J}=\vec{L}+\vec{S}$~.
As an example, in this notation the nucleon and delta wave functions are given by $\left| \, ^{2}8_{1/2} \, [56,0_1^+] \, \right>$ and $\left| \, ^{4}10_{3/2} \, [56,0_1^+] \, \right>$, respectively. 

\section{HQM for baryons}
\label{hQM}

\label{hQM}
In the hQM is supposed that the quark interaction is hypercentral, namely it only depends on the hyperradius \cite{pl,chin},
\begin{equation}
  	V_{3q}(\vec{\rho},\vec{\lambda}) = V(x) \mbox{ },
\end{equation}
where $x = \sqrt{\vec{\rho}^2 + \vec{\lambda}^2}$ is the hyperradius \cite{baf}. Thus, the space part of the three quark wave function, $\psi_{space}$, is factorized as
\begin{equation}
	\psi_{space} = \psi_{3q}(\vec{\rho},\vec{\lambda}) = \psi_{\gamma \nu}(x){Y}_{[{\gamma}]l_{\rho}l_{\lambda}}({\Omega}_{\rho},{\Omega}_{\lambda},\xi) \mbox{ },
\label{psi}
\end{equation}
where the hyperradial wave function, $\psi_{\gamma \nu}(x)$, is labeled by the grand angular quantum number $\gamma$ and the number of nodes $\nu$. ${Y}_{[{\gamma}]l_{\rho}l_{\lambda}}({\Omega}_{\rho},{\Omega}_{\lambda},\xi)$ are the hyperspherical harmonics, with angles ${\Omega}_{\rho}=({\theta}_{\rho},{\phi}_{\rho})$, ${\Omega}_{\lambda}=({\theta}_{\lambda},{\phi}_{\lambda})$ and hyperangle, $\xi = \arctan {\frac{\rho}{\lambda}}$ \cite{baf}.
The dynamics is contained in $\psi_{\gamma \nu}(x)$, which is a solution of the hyperradial equation
\begin{equation}
	\begin{array}{l}
	[\frac{{d}^2}{dx^2}+\frac{5}{x}~\frac{d}{dx}-\frac{\gamma(\gamma+4)}{x^2}] \psi_{\gamma \nu}(x) \\
	\hspace{1cm} = \mbox{ } - 2m~[E-V_{3q}(x)]~~\psi_{\gamma \nu}(x)  \mbox{ }.
	\end{array}
\label{hyrad}
\end{equation}
In the hQM, the quark interaction has the form \cite{pl,chin}
\begin{equation}
	V(x) = -\frac{\tau}{x} + \alpha x \mbox{ },
	\label{h_pot}
\end{equation}
where $\tau$ and $\alpha$ are free parameters, fitted to the reproduction of the experimental data. Eq. (\ref{h_pot}) can be seen as the hypercentral approximation of a Cornell-type quark interaction \cite{Eichten:1978tg}, whose form can be reproduced by Lattice QCD calculations \cite{LQCD}.
Now, to introduce splittings within the $SU(6)$ multiplets, an SU(6)-breaking term must be added. In the case of the hQM, such violation of the $SU(6)$ symmetry is provided by the hyperfine
interaction \cite{deru,ik}. The complete hQM hamiltonian is then \cite{pl,chin}
\begin{equation}
	H_{\rm hQM} = 3m + \frac{\vec{p}_\rho^{~2}}{2m} + \frac{\vec{p}_\lambda^{~2}}{2m}-\frac{\tau}{x} +
	\alpha x + H_{\rm hyp} \mbox{ },
	\label{H_hCQM}
\end{equation}
where $\vec{p}_\rho$ and $\vec{p}_\lambda$ are the momenta conjugated to the Jacobi coordinates $\vec \rho$ and $\vec \lambda$.
In addition to $\tau$ and $\alpha$, there are two more free parameters in the hQM, the constituent quark mass, $m$, and the strength of the hyperfine interaction. 
The former is taken, as usual, as $1/3$ of the nucleon mass. The latter, as in the case of $\tau$ and $\alpha$, is fitted in \cite{pl} to the reproduction of the *** and **** resonances reported in the PDG \cite{Nakamura:2010zzi}.


%

%
\begin{table}
\caption[]{The select strong decay widths of **** nucleon resonances (in MeV) from Ref.\cite{strong2015} . The spectrum is computed using  the {U(7) Model} of Sec. \ref{U(7) algebraic model} and Refs. \cite{Bijker:1994yr-first,Bijker:1994yr} and  {Hypercentral QM} of Sec. \ref{hQM} and Refs. \cite{pl,chin}, in combination with the relativistic phase space factor of Eq. (\ref{eqn:rel-PSF}) and the values of the model parameters of Table \ref{tab:parameters} (second column). The experimental values are taken from Ref. \cite{Nakamura:2010zzi}. Decay channels labeled by -- are below threshold. The symbols ($S$) and ($D$) stand for $S$ and $D$-wave decays, respectively.}
\label{tab:nuc} 
\centering
\scalebox{1}[1]{
\begin{tabular}{cccccccccc}
\hline
\hline
\noalign{\smallskip}
Resonance & Status & $M$ [MeV] & $N \pi$ & $N \eta$ & $\Sigma K$ & $\Lambda K$ & $\Delta \pi$  &  \\
\noalign{\smallskip}
\hline
\noalign{\smallskip}                                                        
$N(1440)P_{11}$          & **** & 1430-1470 & $110-338$ & $0-5$  &      &      &  $22-101$ & Exp. \\
$^28_{1/2}[56,0^+_2]$ &        & 1444          & 85        & --     & --   & --   & 13  &  U(7) \\
$^{2}8_{ 1/2}[56,0_2^+]$ &       & 1550          & 105       & --     & --   & --   & 12  &  hQM \\  \\
$N(1520)D_{13}$ & **** & 1515-1530 & 102       & 0      &      &       & 342 &     Exp. \\
$^28_{3/2}[70,1^-_1]$ &        & 1563 & 134       & 0      & --   & --   & 207  &  U(7) \\ 
$^{2}8_{ 3/2}[70,1_1^-]$ &       & 1525          & 111        & 0     & --   & --   & 206 & hQM  \\  \\
$N(1535)S_{11}$ &**** & 1520-1555 & $44-96$ & $40-91$ & & & $< 2$ &  Exp. \\
$^28_{1/2}[70,1^-_1]$ &        & 1563 &  63 & 75 & -- & --  & 16  &  U(7) \\
$^{2}8_{ 1/2}[70,1_1^-]$ &       & 1525          & 84       & 50    & --   & --   & 6  &  hQM  \\  \\
$N(1650)S_{11}$ & **** & 1640-1680 & $60-162$ & $6-27$ & & $4-20$ & $0-45$ & Exp. \\
$^48_{1/2}[70,1^-_1]$ &        & 1683 & 41 & 72 & -- & 0 & 18 &   U(7) \\ 
$^{2}8_{ 1/2}[70,1_2^-]$ &       & 1574          & 51        & 29     & --   & 0  & 4  &  hQM  \\  \\
$N(1675)D_{15}$ & **** & 1670-1685 & $46-74$ & $0-2$ & & $<2$ & $65-99$ & Exp. \\
$^48_{5/2}[70,1^-_1]$ &        & 1683 & 47 & 11  & -- & 0 & 108 &  U(7) \\ 
$^{4}8_{ 5/2}[70,1_1^-]$ &                   &  1579          & 41        & 9     & --   & --   & 85 &  hQM  \\  \\
$N(1680)F_{15}$ & **** & 1675-1690 & $78-98$ & $0-1$ & & & $6-21$ &  Exp. \\
$^28_{5/2}[56,2^+_1]$ &        & 1737 &   121 & 1 & -- & 0 & 100 &   U(7) \\ 
$^{2}8_{ 5/2}[56,2_1^+]$ &                   &  1798          & 91       & 0     & 0   & 0   & 92  & hQM  \\
\noalign{\smallskip}
\hline
\hline
\end{tabular} }
\end{table}

\section{Two-body strong decays of light nucleon resonances in the $^3P_0$ pair-creation model}
\label{Strong decay widths} 
Here, we present some  our results for the two-body strong decay widths of nucleon resonances  in the $^3P_0$ pair-creation model. 
The decay widths are computed as \cite{Micu,LeYaouanc,Ackleh:1996yt,Barnes:2005pb,Ferretti:2013faa,Ferretti:2013vua,strong2015,bottomonium} 
\begin{equation}
	\Gamma_{A \rightarrow BC} = \Phi_{A \rightarrow BC}(q_0) \sum_{\ell, J} 
	\left| \left\langle BC \vec q_0  \, \ell J \right| T^\dag \left| A \right\rangle \right|^2 \mbox{ },
\end{equation}
where, $\Phi_{A \rightarrow BC}(q_0)$ is the relativistic  phase space factor: 
\begin{equation}
	\label{eqn:rel-PSF}
	\Phi_{A \rightarrow BC}(q_0) = 2 \pi q_0 \frac{E_b(q_0) E_c(q_0)}{M_a}  \mbox{ },
\end{equation}
depending on $q_0$ and on the energies of the two intermediate state hadrons, $E_b = \sqrt{M_b^2 + q_0^2}$ and $E_c = \sqrt{M_c^2 + q_0^2}$.
We assumed harmonic oscillator wave functions, depending on a single oscillator parameter $\alpha_{\rm b}$ for the baryons and $\alpha_{\rm m}$ for the mesons. 
The coupling between the final state hadrons $\left| B \right\rangle$ and $\left| C \right\rangle$ is described in terms of a spherical basis \cite{strong2015}. 
Specifically, the final state $\left| BC \vec q_0  \, \ell J \right\rangle$ can be written as
\begin{equation}
	\begin{array}{rcl}
	\left| BC \vec q_0  \, \ell J \right\rangle & = & \sum_{m,M_b,M_c} 
	\left\langle J_b M_b J_c M_c \right| \left. J_{bc} M_{bc} \right\rangle \\ 
	& & \left\langle J_{bc} M_{bc} \ell m \right. \left| J M \right\rangle \frac{Y_{\ell m}(\hat{q})}{q^2} \delta(q-q_0)  \\
	& & \left| (S_b, L_b) J_b M_b \right\rangle \left| (S_c, L_c) J_c M_c \right\rangle  \mbox{ },
	\end{array} 
\end{equation}
where the ket $\left| BC \vec q_0  \, \ell J \right\rangle$ is characterized by a relative orbital angular momentum $\ell$ between $B$ and $C$ and a total angular momentum $\vec{J} = \vec{J}_b + \vec{J}_c + \vec{\ell}$.

The transition operator of the $^{3}P_0$ model is given by \cite{Ferretti:2013faa,Ferretti:2013vua,strong2015,bottomonium}:
\begin{eqnarray}
\label{eqn:Tdag}
T^{\dagger} &=& -3 \, \gamma_0^{\rm eff} \, \int d \vec{p}_4 \, d \vec{p}_5 \, 
\delta(\vec{p}_4 + \vec{p}_5) \, C_{45} \, F_{45} \,  
{e}^{-r_q^2 (\vec{p}_4 - \vec{p}_5)^2/6 }\, 
\nonumber\\
&& \hspace{0.5cm}  \left[ \chi_{45} \, \times \, {\cal Y}_{1}(\vec{p}_4 - \vec{p}_5) \right]^{(0)}_0 \, 
b_4^{\dagger}(\vec{p}_4) \, d_5^{\dagger}(\vec{p}_5)    \mbox{ }.
\label{3p0}
\end{eqnarray}
Here, $b_4^{\dagger}(\vec{p}_4)$ and $d_5^{\dagger}(\vec{p}_5)$ are the creation operators for a quark and an antiquark with momenta $\vec{p}_4$ and $\vec{p}_5$, respectively.
The $q \bar q$ pair is characterized by a color singlet wave function $C_{45}$, a flavor singlet wave function $F_{45}$, a spin triplet wave function $\chi_{45}$ with spin $S=1$ and a solid spherical harmonic ${\cal Y}_{1}(\vec{p}_4 - \vec{p}_5)$, since the quark and antiquark are in a relative $P$ wave. 
The operator $\gamma_0^{\rm eff}$ of Eq. (\ref{eqn:Tdag}) is an effective pair-creation strength \cite{Ferretti:2013faa,Ferretti:2013vua,strong2015,bottomonium,Kalashnikova:2005ui}, defined as
\begin{equation}
	\label{eqn:gamma0-eff}
	\gamma_0^{\rm eff} = \frac{m_n}{m_i} \mbox{ } \gamma_0 ,
\end{equation}
with $i$ = $n$ (i.e. $u$ or $d$) or $s$ (see Table \ref{tab:parameters}). 

\begin{table}[htbp]  
\begin{center}
\begin{tabular}{ccc} 
\hline 
\hline \\
Parameter           & Value  $U(7)$  & Value  hQM  \\ \\
\hline \\
$\gamma_0$          & 14.3     &     13.319           \\  
$\alpha_{\rm b}$ & 2.99 GeV$^{-1}$  & 2.758 GeV$^{-1}$   \\  
$\alpha_{\rm m}$ & 2.38 GeV$^{-1}$&  2.454  GeV$^{-1}$  \\
$\alpha_{\rm d}$ & 0.52 GeV$^{-1}$  &  0  \\
$m_{\rm n}$               & 0.33 GeV     &       \\
$m_{\rm s}$               & 0.55 GeV     &     \\   \\
\hline 
\hline
\end{tabular}
\end{center}
\caption{Pair-creation model parameters used in the calculations \cite{strong2015}. In the second column are given the parameters used in the calculations with the relativistic phase space factor of Eq. (\ref{eqn:rel-PSF}) for U(7) model, while in the third column those for hQM. The values of the constituent quark masses $m_{\rm n}$ ($n = u,d$) and $m_{\rm s}$ are taken from Refs. \cite{Ferretti:2013faa,Ferretti:2013vua,bottomonium}.}
\label{tab:parameters}  
\end{table}


Finally, the select  results from our study of Ref. \cite{strong2015}, obtained with the values of the model parameters of Table \ref{tab:parameters} (second column) and the relativistic phase space factor of Eq. (\ref{eqn:rel-PSF}), are reported in Table \ref{tab:nuc}.

\section{Discussion and conclusions }
In this contribution, we discussed some recent results for the open-flavor strong decay widths of **** nucleon resonances  within a modified  $^3P_0$ pair-creation model \cite{strong2015}. The baryon spectrum, we needed in our calculation, was predicted within the U(7) algebraic model \cite{Bijker:1994yr} and the hQM  \cite{pl}, developed by Giannini and Santopinto.  

One can observe that the results of Table \ref{tab:nuc} for **** nucleon resonances from Ref. \cite{strong2015} are quite similar for $N\pi$ and $\Delta\pi$ channels, in both the fits we did for the hQM and U(7) model cases. But it is worthwhile noticing  that the parameters of the $^3P_0$ model are quite different, see Table \ref{tab:parameters}. On the contrary, in the case of the $\eta\pi$ channels the predictions are different in the hQM and U(7) model.  

The possibility of using different models to extract the baryon spectrum helps to understand differences between different types of quark models. 
Another step towards a deeper understanding of this type of processes could be an extension of the quark model to include the continuum components in the baryon wave function. Thus,  in a subsequent paper we will focus on threshold effects and the decays of states close to open and hidden-flavor decay thresholds. This procedure will not only have an effect on the widths, but also on the mass values \cite{selfenergies2016}. 

\begin{acknowledgements}
This work was supported in part by INFN, EPOS, Italy and by PAPIIT-DGAPA, Mexico (Grant No. IN107314).
\end{acknowledgements}



\end{document}